\documentclass[a4paper]{PoS}

\usepackage{amsmath}
\usepackage{amssymb}

\newcommand{\eqn}{equation}
\newcommand{\al}{\alpha}

\newcommand{\lb}{\left(}
\newcommand{\rb}{\right)}
\newcommand{\ph}{\hat{p}}
\newcommand{\D}{\mathcal{D}}
\newcommand{\M}{\mathcal{M}}

\newcommand{\nc}{\newcommand}

\nc{\beq}{\begin{equation}}
\nc{\eeq}{\end{equation}}
\nc{\bea}{\begin{eqnarray}}
\nc{\eea}{\end{eqnarray}}
\nc{\nn}{\nonumber}

\nc{\veps}{\varepsilon}
\nc{\eps}{\epsilon}
\nc{\as}{\alpha_s}
\nc{\cd}{\cdot}

\title{An alternative subtraction scheme for NLO QCD calculations using
Nagy-Soper dipoles}

\ShortTitle{An alternative subtraction scheme for NLO QCD calculations using
Nagy-Soper dipoles}

\author{Markus Bach,\ \speaker{Tania Robens}\\%
        IKTP, TU Dresden\\
 E-mail: \email{Markus.Bach1@mailbox.tu-dresden.de}, \phantom{E-mail: }\email{Tania.Robens@tu-dresden.de }
}

\author{Cheng Han Chung\\%
        Supercomputing Research Center, National Cheng Kung University\\
        E-mail: \email{chenghan@mail.ncku.edu.tw}
}

\abstract{In this work, we briefly report on the current status of an alternative subtraction scheme which is based on the splitting kernels of an improved parton shower prescription. Our focus is here on more recent developments as well as generic arguments on the scaling behaviour.
}

\FullConference{11th International Symposium on Radiative Corrections
(Applications of Quantum Field Theory to Phenomenology) \\
                 22-27 September 2013\\
                 Lumley Castle Hotel, Durham, UK}

\begin{document}
\bibliographystyle{unsrt}
\section{Introduction: subtraction schemes in the NLO era}
It is indisputable that higher order corrections are needed to correctly predict fully differential distributions for scattering processes at high precision. The recent discovery of a Higgs boson \cite{Aad:2012tfa,Chatrchyan:2012ufa}, which led to the award of this year's Nobel prize \cite{higgsnobel}, is a more than intriguing example that precision physics is indispensable to a correct theoretical interpretation of experimental data accumulated at present and future colliders. An ideal framework for a detailed comparison between theoretical predictions and experimental findings are Monte Carlo event generators. In addition, especially in recent years many tools have been developed which allow for the (semi-)automated calculation of higher order corrections. These tools provide the virtual contributions, whereas the integration over phase space as well as the calculation of the real emission part is usually performed by a standard Monte Carlo generator, as e.g. Sherpa \cite{Gleisberg:2008ta} or Herwig++ \cite{Bahr:2008pv}. The next-to-leading order (NLO) matching of such higher order calculations to parton showers is equally well understood.

The implementation of NLO calculations into numerical tools exhibits a caveat stemming from the infrared divergence of real and virtual NLO contributions, which originate from different phase spaces: although in the sum of all contributions, the infinite parts exactly cancel, the behaviour of the divergence needs to be parametrized, e.g. by infinitesimal regulators. The implementation of such regulators into numerical codes can result in large unphysical numerical uncertainties. A way to circumvent this problem is the introduction of subtraction schemes, which efficiently reshuffle the divergent terms such that a numerically stable evaluation becomes possible for contributions stemming from both Born-type and real-emission kinematics. We here discuss a specific scheme and its properties, which has first been proposed in \cite{Chung:2010fx}, using splitting kernels as well as mapping prescriptions which were already suggested in the framework of an improved parton shower \cite{Nagy:2007ty,Nagy:2008ns,Nagy:2008eq}. It was further developed for processes with an arbitrary number of final states in \cite{Chung:2012rq}, and a recent review was presented in \cite{Robens:2013wga}. Furthermore, the scheme has recently been implemented in an automated way within the HelacNLO framework \cite{Bevilacqua:2013iha}. We here largely follow the notation of \cite{Chung:2010fx,Chung:2012rq,Robens:2013wga} and only briefly review the setup of the scheme, rather focussing on newer developments which have not been presented previously. We equally reemphasize a theoretical argument \cite{Chung:2012rq} which in principle allows for an even further reduction of the scaling.

\section{Subtraction Schemes}\label{sec:schemes}
Higher order subtraction schemes make use of factorization of the real-emission matrix element in the soft or collinear limits, leading to the decomposition 
$\left|{\cal M}_{m+1}(\hat p)\right|^2 \longrightarrow \D_\ell\,\otimes\,\left| {\cal M}_{m}( p) \right|^2$
 \cite{Altarelli:1977zs,Bassetto:1984ik,Dokshitzer:1991wu}.
Here, $\D_\ell$ are the dipoles which contain the respective singularity structure. The symbol $\otimes$ denotes a correct convolution in colour, spin, and flavour space, and $\hat p/\, p$ are momenta in $(m+1)/\,m$-parton phase space. The subtracted contributions are then given by
\bea
\textstyle
\label{countertermfinite85}
\sigma^{\text{NLO}}&=&\underset{\text{finite}}
{\underbrace{\int_{m+1}\left[
d\sigma^R-d\sigma^A\right]}}+\underset{\text{finite}}
{\underbrace{\int_{m+1}\,d\sigma^A+\int_m\,d\sigma^V}}        
\eea
where
\begin{alignat}{53}
\label{explicitexpressionsNLO}
\textstyle
\int_m\, \left[d\sigma^B\,+\,d\sigma^V\,+\,\int_1\,d\sigma^A\right]& =\int    dPS_m
\left[\left| {\cal M}_{m} \right|^2\,+\,\left| {\cal M}_{m} \right|^2_{\text{one-loop}}\,+\,
\sum_\ell\,\mathcal{V}_\ell\,\otimes\,\left| {\cal M}_{m} \right|^2\right], \notag \\
\int_{m+1}\,\left[ d\sigma^R - d\sigma^A \right]&=\int dPS_{m+1} \left[\left| {\cal M}_{m+1} \right|^2 \,-\, \sum_\ell\,D_\ell\,\otimes\,\left| {\cal M}_{m} \right|^2\right],  
\end{alignat}
and where $\int\,d\,PS$ denotes the integration over the respective phase space, including all symmetry and flux factors.
The symbols $d\sigma^B,\,d\sigma^V,\,d\sigma^R$ stand for the Born, virtual and real-emission contributions of the calculation, while real-emission subtraction terms are summarized as $d\sigma^A$.
Since $\left|{\cal M}_{m+1}\right|^2$ and 
$\left| {\cal M}_{m} \right|^2$ live in different
phase spaces, their momenta need to be mapped via a mapping function. Furthermore, the subtraction term
$\D_\ell$ and its one-parton integrated counterpart $\mathcal{V}_\ell$ are related by 
$\mathcal{V}_\ell\,=\, \int\,d\xi_p \,\D_{\ell},$
where $d\xi_p$ is an unresolved one-parton integration measure. The following ingredients therefore define a subtraction scheme:
(a) a suitable mapping from $(m+1)$ to $m$ parton phase space which guarantees energy-momentum conservation as well as on-shellness, and
(b) an efficient parametrization of the one-parton integration measure $d\xi_p$.
While  the number of reevaluations of the underlying Born matrix element for the real emission subtractions in Eqn. (\ref{explicitexpressionsNLO}) is determined by (a), the complexity of the integrated counterterms depends on (b). Currently, two major schemes for NLO subtraction are on the market, namely
the Catani-Seymour (CS) dipole scheme \cite{Catani:1996vz,Catani:2002hc}, and
the Frixione-Kunszt-Signer (FKS) subtraction \cite{Frixione:1995ms}.
In the scheme discussed here,
we use the splitting kernels of an improved parton shower as a basis for the real emission subtraction terms\footnote{This equally promises to facilitate the matching to the improved parton shower, cf. discussion in \cite{Robens:2013wga} and references therein.}, and
we apply a momentum mapping which leads to an overall scaling behaviour $\sim\,N^2$ for a process with $N$ partons in the final state.
The number of matrix element reevaluations is thereby reduced by a factor proportional to the number of final state particles of the process with respect to the CS scheme.

\section{Nagy-Soper subtraction: Setup and relation to improved parton shower}\label{sec:ns-sub}
We denote
four-momenta in the Born-type kinematics by unhatted quantities $p_i$, while the real emission phase space momenta are denoted by hatted quantities $\ph_i$;
initial state momenta are labelled $p_a$ and $p_b$, where $Q\,=\,p_a+p_b$ and with $Q^2$ being the squared centre-of-mass energy, with equivalent relations in the real emission phase space;
generally, $\ph_\ell$ labels the emitter, $\ph_j$ the emitted parton and $\ph_k$ the spectator. 

\subsection{Scheme setup}
The scheme discussed here uses the splitting kernels of an improved parton shower \cite{Nagy:2007ty,Nagy:2008ns,Nagy:2008eq} as a basis for the subtraction terms. 
We can therefore write \cite{Nagy:2007ty}
\beq
\label{QCDFactorizationm1tVm}
\mid {\cal M}_\ell(\{\hat p, \hat f\}_{m+1})\rangle\,=\,t^\dagger_\ell(f_\ell \to \hat f_\ell + 
\hat f_{j})\,V^\dagger_\ell(\{\hat p, \hat f\}_{m+1})\,\mid {\cal M}(\{ p,  f\}_{m})\rangle.
\eeq
Here,  $\mid {\cal M}_\ell(\{\hat p, \hat f\}_{m+1})\rangle$ and $\mid{\cal M}(\{ p,  f\}_{m})\rangle$ denote the matrix elements in real emission \mbox{($m+1$)} and Born-type ($m$) phase space and $V_\ell,\,t_\ell$ the factorization operators in colour and spin space. For fermionic emitters, the splitting functions are diagonal in helicity space; therefore, the real emission subtraction terms are directly given by the spin averaged functions
${\textstyle \overline{W}_{\ell\,\ell}\,=\,\frac{1}{2}v^2_\ell,}$
with $v_\ell^2$ being defined by Eqn. (43) in \cite{Chung:2010fx}. 
In case of gluonic emitters, information on the gluon polarization needs to be retained, and soft/ collinear divergences from interference terms arise.
In our scheme, these are treated using dipole partitioning functions $A_{\ell k}$ \cite{Nagy:2008eq}, which redistribute the singularities to contributions $W^{(\ell)}_{\ell k},\,W^{(k)}_{\ell k}$, where $p_\ell,\,p_k$ take over the kinematic role of the mother parton in the mapping, respectively. The subtraction term is then split into a purely collinear and a soft/ collinear part
$\langle \nu'| W_{\ell\ell}- W_{\ell k}|\nu\rangle \,=\,\langle \nu'| \left( {W}_{\ell\ell} - {W}_{\ell\ell}^{\textrm {eik}} \right)
+ \left({W}_{\ell\ell}^{\textrm {eik}}  - {W}_{\ell k}\right)|\nu\rangle,$
where $|\nu\rangle,\,|\nu'\rangle$ denote the gluon polarization of the mother parton connected to the Born-type matrix element in Eqn. (\ref{QCDFactorizationm1tVm}). Our specific choice of the dipole partitioning functions leads to
\beq
\label{interferencespinaveragedsplittingfunction}
\Delta W_{\ell k} \,=\, \overline{W}_{\ell\ell}^{\textrm {eik}}  - \overline{W}_{\ell k} \,=\,4\,\pi\,\al_s
\frac{2\, (\hat p_\ell\cdot\hat p_k)\, (\hat p_\ell\cdot\hat Q) } 
{(\hat p_\ell\cdot\hat p_j)\,
\left[(\hat p_j\cdot\hat p_k)\, (\hat p_\ell\cdot\hat Q)+(\hat p_\ell\cdot\hat p_j)\,(\hat p_k\cdot\hat Q) \right]}
\eeq
where $\overline{W}_{\ell\ell}^{\textrm {eik}}$ is the spin-averaged eikonal factor.
All quantities are defined as in \cite{Chung:2010fx,Chung:2012rq,Robens:2013wga}.
\subsection{Final state momentum mapping and scaling behaviour}\label{sec:mapscal}
The improved scaling behaviour of our scheme mainly results from the specific mapping between the real emission and Born-type kinematic phase spaces for final state emitters. For final state mappings, we use the whole remainder of the event as a spectator in terms of momentum redistributions; therefore, we have 
\begin{eqnarray}
\label{eq:fin_map}
p_\ell\,=\,\frac{1}{\lambda_\ell}\,(\hat p_\ell + \hat p_j)-\frac{1 - \lambda_\ell + y_\ell}{2\, \lambda_\ell\, a_\ell}\, Q,\;\;
p_n^\mu\,=\,\Lambda (K,\hat{K})^\mu{}_\nu \,\hat p^{\nu}_{n} ,\quad n\notin\{\ell,j=m+1\},
\end{eqnarray}
with
${\textstyle \Lambda(K,\hat K)^{\mu}_{\;\;\nu} \,=\,g^{\mu}_{\;\;\nu}\,-\,\frac{2\,( K+\hat K)^{\mu}\,(K+\hat K)_{\nu}}{(K+\hat K)^{2}}\,
+\,\frac{2\,{K}^{\mu}\,\hat{K}_{\nu}}{\hat K^{2}}\,},$
where
${\textstyle
y_\ell = \frac{P_\ell^2}{2\, P_\ell\cdot Q - P_\ell^2}.}$
We furthermore introduced
${\textstyle \lambda_\ell\lb y_\ell,a_\ell \rb\,=\,\sqrt{\left(1+y_\ell\right)^2-4\,a_\ell\,y_\ell},}\;$ ${\textstyle K\,=\, Q - p_\ell,}\;$ ${\hat{K}\,=\, Q - P_\ell,\,a_\ell\lb P_\ell,Q\rb\,=\,\frac{Q^2}{2\, P_\ell\,\cdot\,Q-P_\ell^2}}$, with $P_\ell\,=\,\ph_\ell+\ph_j$. 
Note that it is the {\sl global} mapping for all remaining particles in Eqn. (\ref{eq:fin_map}) that is responsible for the reduced number of Born-type matrix reevaluations. 
For the real emission subtraction terms, we then obtain the total contribution
\begin{\eqn}\label{eq:master_sub}
d\sigma^{A}_{ab}(\ph_a,\ph_b)\,=\,d\sigma^{A,a}_{ab}(\ph_a,\ph_b)+d\sigma^{A,b}_{ab}(\ph_a,\ph_b)
+\sum_{\ell\,\neq\,a,\,b} d\sigma^{A,\ell}_{ab}(\ph_a,\ph_b),
\end{\eqn}
with the sum over all possible final state emitters. For a specific emitter $\ph_\ell$, it is explicitly given by
\begin{eqnarray}\label{eq:counter_fin}
d\sigma^{A,\ell}_{ab}(\hat{p}_a,\hat{p}_b)&=&\frac{N_{m+1}}{\Phi_{m+1}}
\sum_{j}\Bigg\{
\left[\D_{gqq}(\ph_{j})\delta_{g;q,q_{j}}+\D_{ggg}(\ph_{j})
\delta_{g;g,g_{j}}\,\right]|\M_{\text{Born},g}|^{2}(p_{a},p_{b};p_{n})\nn\\
&&\,+\left[ \D_{qqg}(\ph_{j})\delta_{q;g,q_{j}}\,+\,\D_{qqg}(\ph_{j})
\delta_{q;q,g_{j}}\right]|\M_{\text{Born},q}|^{2}(p_{a},p_{b};p_{n})\Bigg\},
\end{eqnarray}
where $\Phi_{m+1}$ denotes the flux factor\footnote{The $\delta_{f_\ell;\hat{f}_\ell\,\hat{f}_j}$ functions ensure the existence of the respective splittings $f_\ell\,\rightarrow\,\hat{f}_\ell\,\hat{f}_j$ in flavour space.}.
The subtraction terms can be split into collinear and interference parts:
 \begin{\eqn}\label{eq:delljtot}
\D_{f_\ell \hat{f}_\ell \hat{f}_{j}}(\ph_\ell, \ph_j)\,=\,\D^{\text{coll}}_{f_\ell \hat{f}_\ell \hat{f}_{j}}(\ph_\ell, \ph_j)\,+\,\delta_{\hat{f}_{j},g}\sum_{k\,\neq\,(\ell,j)}\D^{\text{if}}(\ph_\ell,\ph_j,\ph_{k}),
\end{\eqn}
where $\D^{\text{if}}(\ph_\ell,\ph_j,\ph_{k})$ denotes an interference contribution with $\ph_{k}$ acting as a spectator. 
Each of the contributions in Eqn. (\ref{eq:delljtot}) requires exactly {\sl one} global mapping, i.e. the number of mappings and thereby matrix reevaluations in the real emission subtraction terms behaves like $\sim\,\#(\ell j)$. This leads to an overall scaling behaviour $\sim\,N^2/2$, where $N$ is the number of final state particles, which can in principle be improved even further. In \cite{Frederix:2009yq}, the authors show that within the MadFKS environment, a constant scaling behaviour can be achieved; i.e., for certain types of processes, the number of reevaluations of underlying Born-type matrix elements in the real-emission subtraction terms remains constant. This relies on the fact that any $m+1$ phase space can be decomposed into disjoint partitions that are specified by their behaviour for one of the partons $\ph_i$ becoming soft or collinear to at most one other parton $\ph_j$,
where the sum of all FKS partitions reproduces the whole phase space\footnote{Note that the notation between \cite{Frederix:2009yq} and this work differs in the fact that in \cite{Frederix:2009yq}, $\ph_i$ labels the emitted parton that becomes soft or collinear, while in our case this parton is denoted by $\ph_j$. For sake of consistency, we stick to the notation proposed in \cite{Frederix:2009yq} in the above discussion.}, such that
$\sum_{(i,j)\in \mathcal{P}_\text{FKS}}\,\mathcal{S}_{ij}\,=\,1.$ Here, $\mathcal{P}_\text{FKS}$ is the set of FKS pairs labelled by the parton indices $(i,j)$ , and $\mathcal{S}_{ij}$ are the $S$-functions \cite{Frederix:2009yq} which project out the respective FKS partition $(i,j)$.
 Furthermore, several partitions render exactly the same contribution to the final observable, and therefore the evaluation of only one of these is sufficient:
\begin{\eqn}\label{eq:fks_sym}
d\sigma^{(n+1)}(r)\,=\,\sum_{(i,j)\in\overline{\mathcal{P}}_\text{FKS}}\xi_{ij}^{(n+1)}(r)d\sigma_{ij}^{(n+1)}(r),
\end{\eqn}
where $\xi_{ij}^{(n+1)}(r)$ is the process-dependent symmetry factor that relates the total cross section to the one evaluated in the partition $(i,j)$, and $\overline{\mathcal{P}}_\text{FKS}$ now signifies the set of all nonredundant FKS pairs. 
In the scheme discussed here, the subtraction term that reflects the divergences of $\mathcal{S}_{ij}$ is given by Eqn. (\ref{eq:delljtot}), such that
all contributions from the soft/ collinear divergence of $\ph_i,\,\ph_k$ are transferred to the interference term $\mathcal{D}^\text{if}(\ph_k,\ph_i,\,\ph_j)$, corresponding to the singularity structure of a {\sl different} partition, namely $\mathcal{S}_{ik}$. All terms in Eqn. (\ref{eq:delljtot}) come with the same mapping, and, as in the FKS prescription in \cite{Frederix:2009yq}, only the set of nonredundant contributions needs to be evaluated, all others being related by symmetry. Increasing the number of final state gluons then leads to a change in the constant $\xi_{ij}^{(n+1)}(r)$ but does not call for the evaluation of a larger number of nonredundant contributions, as the number of elements in $\overline{\mathcal{P}}_\text{FKS}$ remains constant. Therefore, following this prescription, our scheme equally exhibits a constant scaling behaviour, when the number of gluons in the real emission final state is increased.
\subsection{Subtraction terms and integrated counterterms}
We devote this subsection to a more detailed discussion of one of the leftover finite parts in the integrated counterterms that is currently evaluated numerically. The existence of finite remainders in these terms is a direct consequence of the modified mapping which leads to the improved scaling behaviour discussed above. Although this constitutes a slight modification with respect to standard schemes such as CS and FKS, it poses no impediment for the implementation of our scheme. 

We here present an example of an approximation for one of these integrals. For this, we focus on an integral which appears in the final state $qqg$ splitting, namely
\begin{\eqn}\label{eq:i3}
I_3(a_\ell)\,=\,-\int^{y_\text{max}}_{0}\,dy\,\left[\frac{(\lambda-1+y)^{2}}{4\,y}+1\right]\,\frac{(1+y)\ln\,x_{0}}{\lambda},
\end{\eqn}
where $\lambda\,\equiv\,\lambda_\ell(y,a_\ell),\;x_0(y,a_\ell)\,=\,\frac{1-\lambda+y}{1+\lambda+y},\,y_\text{max}(a_\ell)\,=\,\lb \sqrt{a_\ell}-\sqrt{a_\ell-1} \rb^2$, and all other parameters are defined in Section \ref{sec:mapscal}. For $a_\ell\,=\,1$, the integral is given by $I_3(1)=\frac{\pi^2}{3}-1$; for all other cases, we use an approximation. We easily find the approximating functions
\begin{eqnarray}
\lefteqn{I_3^{\text{approx},a_\ell\,\leq\,2}(a_\ell)\,=\,
\frac{a_\ell}{4}+(1+a_\ell^2)\left[\frac{\pi^2}{6}-\mathrm{Li}_2(1-y_\text{max})\right]+2\,(a_\ell-1)^2\,\ln(1-y_\text{max})}\nonumber\\
&&+\frac{1}{16}\bigg[ 13\,y_\text{max}(1-a_\ell)^2-5\,a_\ell^2\,y_\text{max}
+\ln\,y_\text{max} \left[ 5
+4\,\lb a_\ell\,
+2\,y_\text{max}\,-3\,a_\ell\,y_\text{max}\rb\right]-15\bigg]\nonumber\\
&&-\frac{(a_\ell-1)\,(a_\ell^2-2.75128\,a_\ell+1.74026)}{0.248486\,a_\ell^3+7.10958\,a_\ell^2-3.11175\,a_\ell-3.68862}\label{eq:i3_1}\\
&&\nonumber\\
\lefteqn{I_3^{\text{approx},2\,\leq\,a_\ell\,\leq\,20}(a_\ell)\,=\,
0.12842+\,\frac{0.560353}{(a_\ell-0.578142)^{0.974664}}}\nonumber\\
&&+\,\frac{a_\ell^4-28.1242\,a_\ell^3+240.479\,a_\ell^2-758.744\,a_\ell+778.706}{1457.74\,a_\ell^4+16001.5\,a_\ell^3-48392.7\,a_\ell^2+64199\,a_\ell-39009.3} \label{eq:i3_2}
\end{eqnarray}
For $a_\ell\,\geq\,20$, a similar approximation applies.
\begin{center}
\begin{figure}
\begin{center}
\begin{minipage}{0.45\textwidth}
\begin{center}
\includegraphics[width=\textwidth]{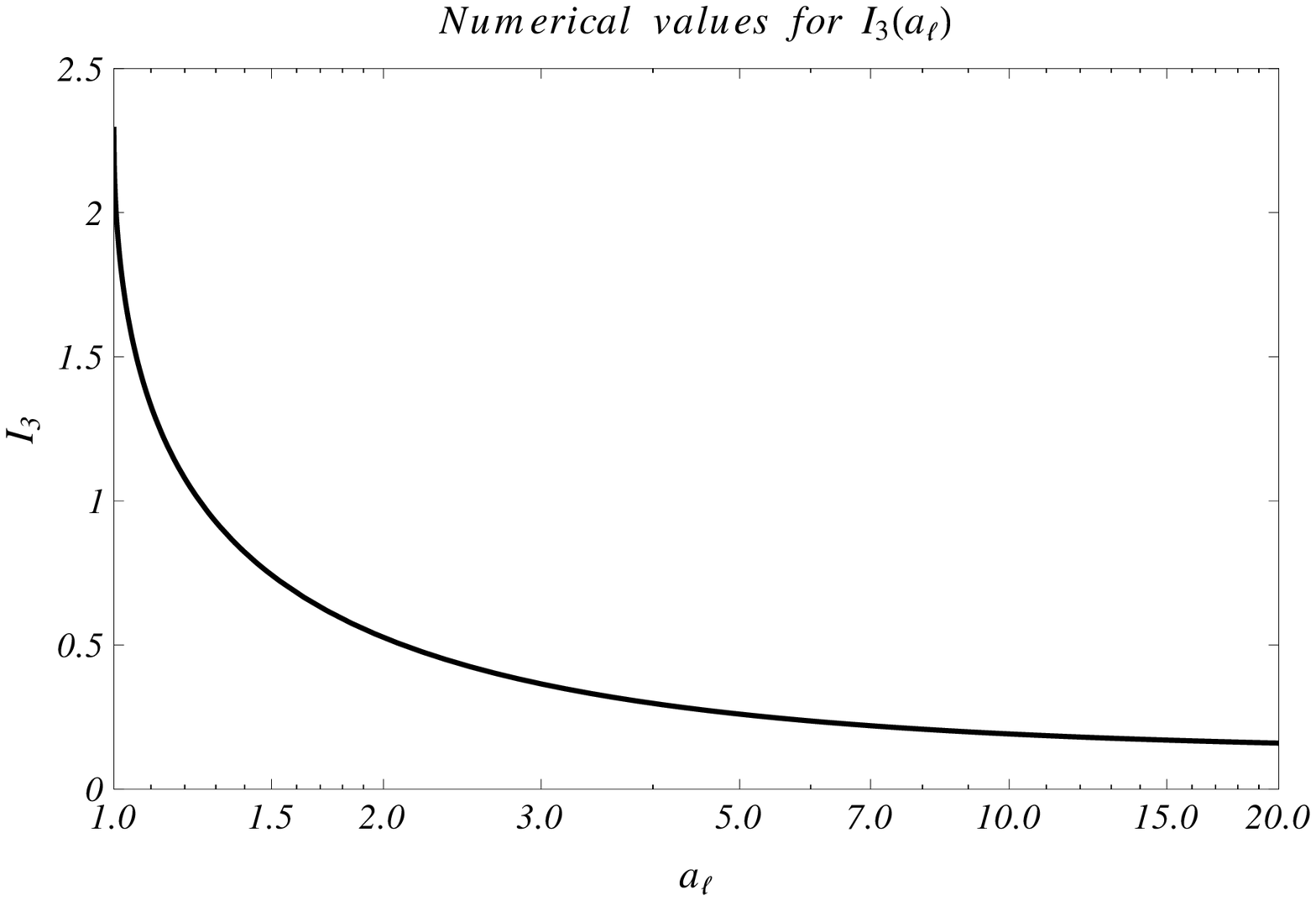}
\end{center}
\end{minipage}
\hspace{2mm}
\begin{minipage}{0.45\textwidth}
\begin{center}
\includegraphics[width=\textwidth]{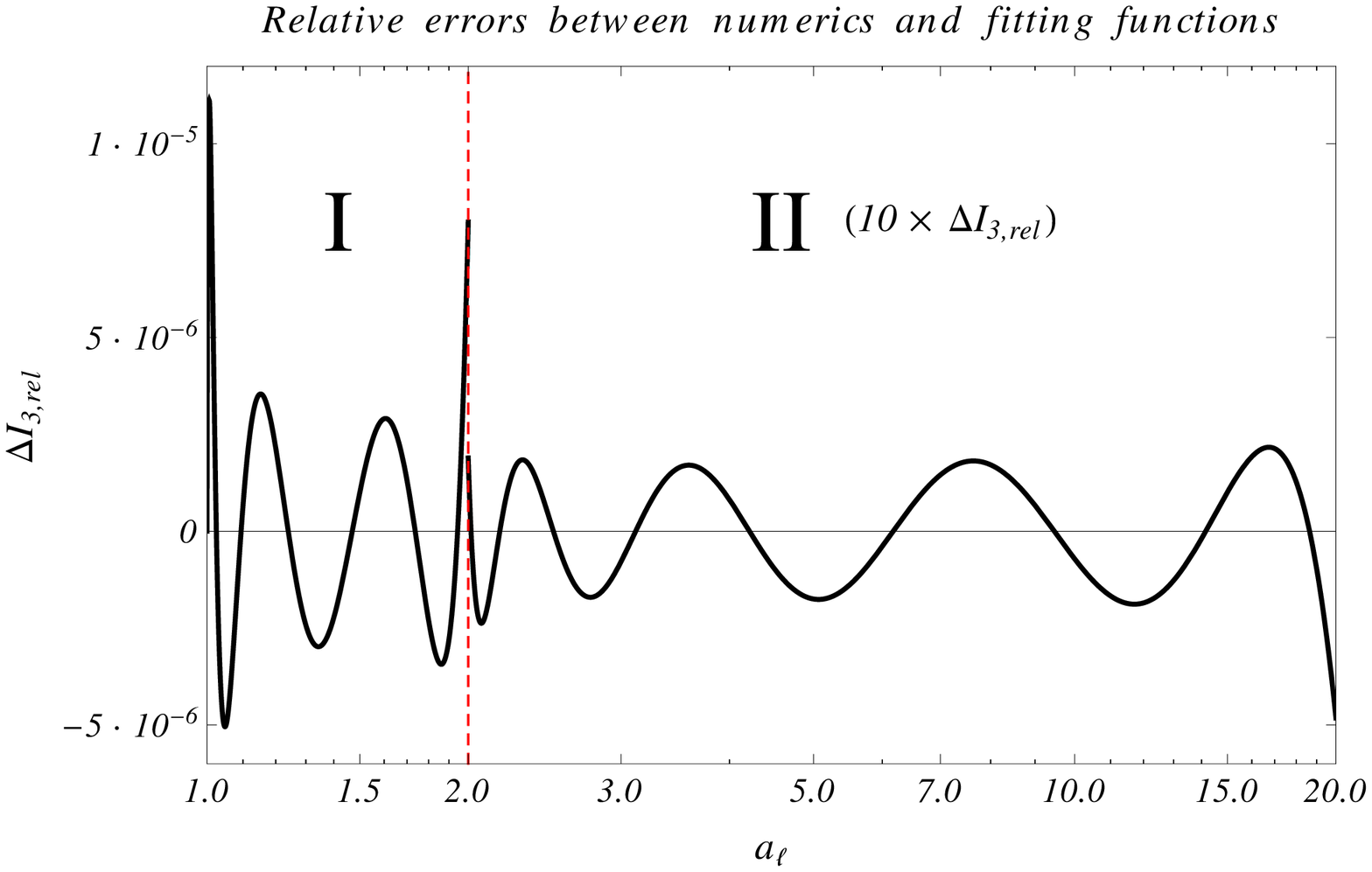}
\end{center}
\end{minipage}
\caption{\label{fig:i3approx} {\sl Left:} Numerical values for $I_3(a_\ell)$ (Eqn. (3.8)), in the range $a_\ell\,\in\,[1,20]$.  {\sl Right:} Relative error between numerical value for $I_3(a_\ell)$ and fitting functions for $a_\ell\,\in\,[1,2]$ (section I) and $a_\ell\,\in\,[2,20]$ (section II), given by Eqns. (3.9) and (3.10) respectively. We find the relative errors are $\mathcal{O}\lb 10^{-5} \rb$, thereby well below typical phase space integration errors. Note that in the above plot the relative error in section II has been multiplied by 10 for better visibility.}
\end{center}
\end{figure}
\end{center}

Figure \ref{fig:i3approx} shows the behaviour of the integral for the range $a_\ell\,\in\,\left[1, 20\right]$, as well as the relative errors between the approximation as given in Eqns. (\ref{eq:i3_1}), (\ref{eq:i3_2}) and the numerically evaluated integral; we found that the errors are $\mathcal{O}\lb 10^{-5} \rb$ for $a_\ell\,\in\,[1,2]$, and an order of magnitude smaller for $a_\ell\,\geq\,2$. We want to emphasize that these relative errors are small compared to the errors typically obtained from the Monte Carlo integration over phase space.
\section{Results}
As an example, we here present results for the process $e^+\,e^-\,\rightarrow\,\text{3 jets}$ \cite{Chung:2012rq}.
The leading order contribution is given by
$e^+\,e^-\,\rightarrow\,q\,\bar{q}\,g,$
and we include virtual corrections, as well as real emission processes
$e^+\,e^-\,\rightarrow\,q\,\bar{q}\,q\,\bar{q},\;e^+\,e^-\,\rightarrow\,q\,\bar{q}\,g\,g$
 \cite{Ellis:1980wv}. The above real emission contributions call for $(8+10)$ matrix element reevaluations per phase space point in the CS and $(4+5)$ reevaluations in our scheme, respectively. 
We display our results in terms of the C distribution \cite{Ellis:1980wv}
${\textstyle C^{(n)}\,=\,3\,\left\{ 1-\sum_{i,j\,=\,1,\,i<j}^n\,\frac{s_{ij}^2}{(2\,p_i\cdot\,Q)\,(2\,p_j\cdot\,Q)}  \right\},\,{(s_{ij}\,=\,2\,p_i \cdot p_j)}}$,
which fulfills all requirements of a jet observable and is infrared finite on the integration level \cite{Catani:1997xc}. 
\begin{center}
\begin{figure}
\begin{center}
\begin{minipage}{0.45\textwidth}
\begin{center}
\includegraphics[width=\textwidth]{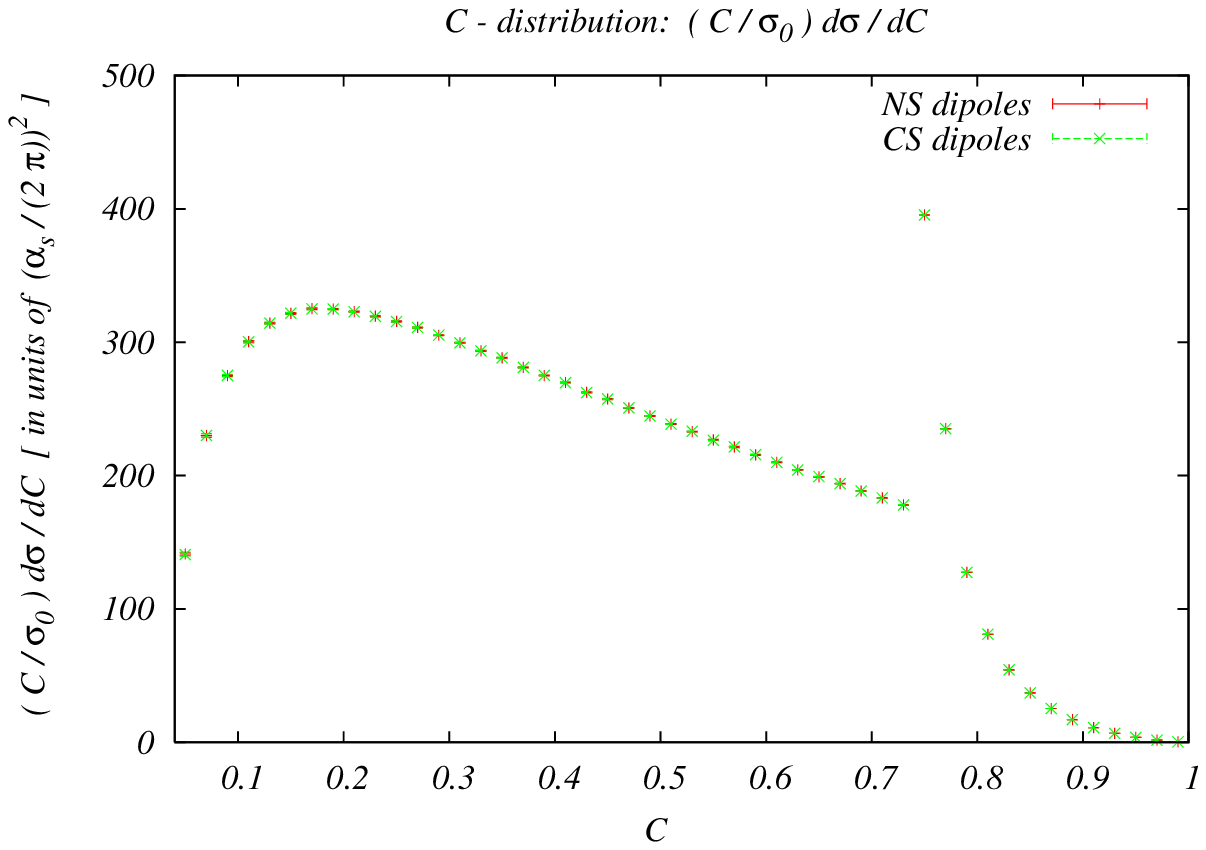}
\end{center}
\end{minipage}
\hspace{2mm}
\begin{minipage}{0.45\textwidth}
\begin{center}
\includegraphics[width=\textwidth]{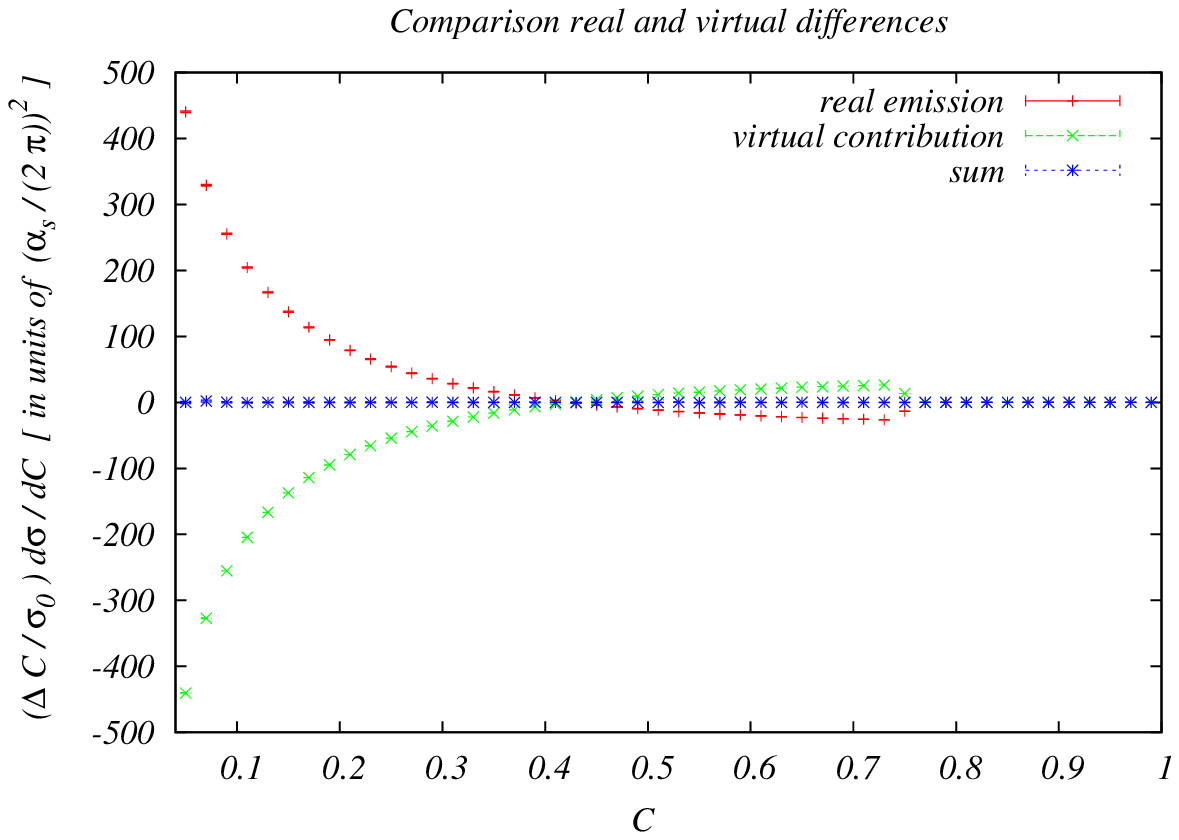}
\end{center}
\end{minipage}
\caption{\label{fig:tot_diff} {\sl Left:} Total result for differential distribution $\frac{C}{\sigma_0}\,\frac{d\sigma^\text{NLO}}{dC}$ using both our scheme (labelled NS, red) and CS (green) dipoles. The standard literature result obtained using the CS scheme is completely reproduced with the NS dipoles. {\sl Right:} {\sl Differences} $\Delta_\text{CS-NS}$ for real emission (red, upper) and virtual (green, lower) contributions, showing that especially for low $C$ values the contributions in the two schemes significantly differ. Adding up $\Delta^\text{real}+\Delta^\text{virt}$ gives 0 as expected.}
\end{center}
\end{figure}
\end{center}
Figure \ref{fig:tot_diff} shows that we reproduce the literature result, numerically obtained from \cite{mike}, and equally found agreement between implementations of both schemes. We want to emphasize that this is indeed a non-trivial statement, since the {\sl differences} between the two schemes for both subtracted real emission as well as virtual contributions are sizeable; therefore, agreement between the two schemes on the per mil level \cite{Chung:2012rq} constitutes a non-trivial validation of our scheme. 
\section{Summary}

We  here reported on some recent progress in the development of an alternative NLO subtraction scheme for QCD calculations, which uses the splitting functions of an improved parton shower as subtraction kernels. We have briefly discussed the setup, and especially the features leading to an improved scaling behaviour of our scheme with respect to one of the standard subtraction schemes. We focussed on possible further improvements of this scaling behaviour, along the lines of a proposal which has first been investigated within the MadFKS framework. Results for the process $e^+\,e^-\,\rightarrow\,3$ jets as well as an example of an analytic approximation for one of the functions which is currently evaluated numerically have been presented. Summarizing, we regard the scheme discussed here as a viable alternative to both CS and FKS subtraction. Our scheme exhibits a smaller number of subtraction terms with respect to CS, and does not call for a reparametrization of the phase space for each emitter/ emitted parton pair, as needed in the FKS scheme. We believe that this direction is worthwhile to investigate, and plan to implement this into a general purpose Monte Carlo event generator in the near future.

\bibliography{NLO_subtraction}


\end{document}